\index{}

\documentclass{ws-mpla}

\newcommand{\ffat}[1]{\mbox {\boldmath $#1$}}

\begin{document}

\markboth{Authors' Names}
{A Formulation without Partial Wave Decomposition}

\catchline{}{}{}{}{}

\title{A FORMULATION WITHOUT PARTIAL WAVE DECOMPOSITION\\
FOR SCATTERING OF SPIN-$\frac{1}{2}$ AND SPIN-$0$ PARTICLES}

\author{\footnotesize I. ABDULRAHMAN}

\address{Departemen Fisika, Universitas Indonesia, Depok 16424, Indonesia}

\author{I. FACHRUDDIN}

\address{Departemen Fisika, Universitas Indonesia, Depok 16424, Indonesia}

\maketitle

\pub{Received (Day Month Year)}{Revised (Day Month Year)}

\begin{abstract}
A new technique has been developed to calculate scattering of spin-$\frac{1}{2}$ and 
spin-$0$ particles. The so called momentum-helicity basis states are 
constructed from the helicity and the momentum states, which are not expanded 
in the angular momentum states. Thus, all angular momentum states are 
taken into account. Compared with the partial-wave approach this technique 
will then give more benefit especially in calculations for higher energies. 
Taking as input a simple spin-orbit potential, the Lippman-Schwinger 
equations for the T-matrix elements are solved and some observables are calculated. 

\end{abstract}

\ccode{PACS Nos.: 21.45.-v}

\section{Introduction}	

As an alternative to the partial wave (PW) decomposition a so called three-dimensional 
(3D) technique has been developed for the nucleon-nucleon (NN) system. It has been 
successfully applied to NN scattering and the deuteron in Refs.~\refcite{nn} and 
\refcite{deuteron}, respectively for the Bonn-B\cite{bonnb} and the AV18\cite{av18} 
realistic potential models. The basic idea in 3D approach is to take 
the momentum states, as part of the basis states, without expanding them 
into the angular momentum states. Thus, all angular momentum states are taken into 
account. It is then very important especially for higher energy region, 
while in PW calculations adding more higher angular momentum states may at some point 
become not feasible, for instance, due to restriction on computer resources. For lower 
energy regions still the 3D technique appears as a good alternative. 

In this work we derived a 3D formulation for scattering of spin-$\frac{1}{2}$ and spin-$0$ 
particles. The work will then be very useful, for example, for kaon-nucleon (KN) 
investigation. The formulation described in Section~\ref{formulation} is simpler 
than that for NN scattering, since our system consists of a pair of nonidentical 
particles with the total spin being $\frac{1}{2}$. There is no antisymmetrizing like 
in Ref.~\refcite{nn}. Finally by means of symmetry relations we get only 
a single Lippmann-Schwinger equation. In Section~\ref{example} we show some 
calculations for a simple spin-orbit interaction based on the Malfliet-Tjon\cite{mt} 
potential. We summarize in Section~\ref{summary}. 

\section{\label{formulation}Formulation}

\subsection{The momentum-helicity basis}

The momentum-helicity basis states are eigenstates to the parity operator $\mathcal{P}$ and the helicity operator $\mathbf{s} \cdot \mathbf{\hat{p}} = \frac{1}{2}\ffat{\sigma} \cdot \mathbf{\hat{p}}$, and are defined as 
\begin{equation}
\bigl|\mathbf{p};\mathbf{\hat{p}\lambda}\bigr>_{\pi} 
    \equiv \frac{1}{\sqrt{2}} \bigl( 1 + \eta_{\pi} \mathcal{P}\bigr) \bigl|\mathbf{p};\mathbf{\hat{p}\lambda}\bigr>,
\label{mhbs}
\end{equation}
with 
\begin{equation}
\bigl|\mathbf{p};\mathbf{\hat{p}\lambda}\bigr> 
    \equiv \bigl|\mathbf{p}\bigr> \bigl|\mathbf{\hat{p}\lambda}\bigr>,
\end{equation}
$\mathbf{p}$ the relative momentum, $\bigl|\mathbf{\hat{p}\lambda}\bigr>$ and $\lambda = \pm \frac{1}{2}$ the helicity eigenstate and eigenvalue for spin $\frac{1}{2}$, $\eta_{\pi} = \pm 1$ the parity eigenvalue, the subscript $\pi = \pm$ the parity eigenstate label. The momentum-helicity basis states in Eq.~(\ref{mhbs}) has the normalization as 
\begin{equation}
\bigl. \bigr._{\pi'} \bigl<\mathbf{p'};\mathbf{\hat{p'}\lambda'}\bigr. \bigl|\mathbf{p};\mathbf{\hat{p}\lambda}\bigr>_{\pi} = \delta_{\eta_{\pi'}\eta_{\pi}} 
            \left[ \delta\bigl(\mathbf{p'} - \mathbf{p}\bigr) \delta_{\lambda'  \lambda} - i \eta_{\pi} 
                   \delta\bigl(\mathbf{p'} + \mathbf{p}\bigr) \delta_{\lambda' -\lambda} \right]
\end{equation}
and the completeness relation as 
\begin{equation}
\frac{1}{2} \sum_{\pi \lambda} \int d\mathbf{p} \bigl|\mathbf{p};\mathbf{\hat{p}\lambda}\bigr>_{\pi} 
\bigl. \bigr. _{\pi} \bigl<\mathbf{p};\mathbf{\hat{p}\lambda}\bigr| = 1 .
\end{equation}

\subsection{The potential and T-matrix elements}

The potential $V$ matrix elements and the T-matrix elements in the momentum-helicity basis are defined as
\begin{eqnarray}
V^{\pi}_{\lambda' \lambda}\left( \mathbf{p'},\mathbf{p} \right) & \equiv & \bigl. \bigr._{\pi} \bigl<\mathbf{p'};\mathbf{\hat{p'}\lambda'}\bigr| V \bigl|\mathbf{p};\mathbf{\hat{p}\lambda}\bigr>_{\pi} \\
T^{\pi}_{\lambda' \lambda}\left( \mathbf{p'},\mathbf{p} \right) & \equiv & \bigl. \bigr._{\pi} \bigl<\mathbf{p'};\mathbf{\hat{p'}\lambda'}\bigr| T \bigl|\mathbf{p};\mathbf{\hat{p}\lambda}\bigr>_{\pi} .
\end{eqnarray}
The potential matrix elements obey the following symmetry relations:
\begin{eqnarray}
V^{\pi}_{\lambda' -\lambda}\left( \mathbf{p'},\mathbf{p} \right) & = & - i \eta_{\pi} V^{\pi}_{\lambda' \lambda}\left( \mathbf{p'},-\mathbf{p} \right) \label{stme1}\\
V^{\pi}_{-\lambda' \lambda}\left( \mathbf{p'},\mathbf{p} \right) & = &  i \eta_{\pi} V^{\pi}_{\lambda' \lambda}\left( -\mathbf{p'},\mathbf{p} \right) \label{stme2}\\
V^{\pi}_{-\lambda' -\lambda}\left( \mathbf{p'},\mathbf{p} \right) & = & V^{\pi}_{\lambda' \lambda}\left( \mathbf{p'},\mathbf{p} \right) 
\end{eqnarray}
and similarly the T-matrix elements. Since the basis states are eigenstates to the helicity operator, it would be convenient if the potential in momentum representation is expressed, in general, as 
\begin{equation}
V\left( \mathbf{p'}, \mathbf{p}\right) \equiv \bigl< \mathbf{p'} \bigl| V \bigr| \mathbf{p} \bigr> 
                                       = \sum_{i} f_{i}\left(p',p,\mathbf{\hat{p'}} \cdot \mathbf{\hat{p}} \right) 
                                          \bigl( \ffat{\sigma} \cdot \mathbf{\hat{p'}} \bigr)^{a_{i}} 
                                          \bigl( \ffat{\sigma} \cdot \mathbf{\hat{p}} \bigr)^{b_{i}} , \label{genpot}
\end{equation}
where $f_{i}\left(p',p,\mathbf{\hat{p'}} \cdot \mathbf{\hat{p}} \right)$ are spin-independent functions. Any spin-dependent potential can be expressed in the form given in Eq.~(\ref{genpot}). 

It follows that for $\mathbf{\hat{p}} = \mathbf{\hat{z}}$ the azimuthal dependence of the potential matrix elements can be separated as
\begin{equation}
V^{\pi}_{\lambda' \lambda}\left( \mathbf{p'},p \mathbf{\hat{z}} \right) = 
e^{i \lambda \phi'} V^{\pi}_{\lambda' \lambda}\left( p',p,x' \right) , \; 
\left( x' \equiv \cos \theta' \right) . \label{azimuth}
\end{equation}
The relation in Eq.~(\ref{azimuth}) applies also to $T^{\pi}_{\lambda' \lambda}\left( \mathbf{p'},p \mathbf{\hat{z}} \right)$. The symmetry relations in Eqs.~(\ref{stme1}) \& (\ref{stme2}) together with the relation in Eq.~(\ref{azimuth}), all for both the potential and the T-matrix elements, lead to the following uncoupled integral equation:
\begin{eqnarray}
\lefteqn{T^{\pi}_{\lambda' \lambda}\left( p',p,x' \right)} && \nonumber\\ 
&& \qquad = V^{\pi}_{\lambda' \lambda}\left( p',p,x' \right) \nonumber \\
&& \qquad \quad + \lim_{\epsilon \rightarrow 0} \int_{0}^{\infty} dp'' \, p''^{2} 
\int_{-1}^{1} dx'' \, 
\frac{\mathcal{V}_{\lambda' \frac{1}{2}}^{\pi \lambda}\left( p',p'',x',x'' \right) 
T^{\pi}_{\frac{1}{2} \lambda}\left( p'',p,x'' \right)}
{\frac{p^{2}}{2 \mu}+ i \epsilon - \frac{p''^{2}}{2 \mu}} , \label{lse}
\end{eqnarray}
which we just call the Lippmann-Schwinger (LS) equation for $T^{\pi}_{\lambda' \lambda}\left( p',p,x' \right)$, with $\mu$ being the reduced mass of the system and 
\begin{equation}
\mathcal{V}_{\lambda' \frac{1}{2}}^{\pi \lambda}\left( p',p'',x',x'' \right) \equiv 
\int_{0}^{2 \pi} d\phi'' \, e^{i \lambda\left( \phi'' - \phi'\right)} 
V^{\pi}_{\lambda' \frac{1}{2}}\left( \mathbf{p'},\mathbf{p''} \right) .
\end{equation}

For $T^{\pi}_{\lambda' \lambda}\left( p',p,x' \right)$ there are symmetry relations given as 
\begin{eqnarray}
T^{\pi}_{-\lambda' \lambda}\left( p',p,x' \right) = -2 \lambda \eta_{\pi} T^{\pi}_{\lambda' \lambda}\left( p',p,-x' \right) \label{symt1}\\
T^{\pi}_{\lambda' -\lambda}\left( p',p,x' \right) = 2 \lambda' \eta_{\pi} T^{\pi}_{\lambda' \lambda}\left( p',p,-x' \right) \label{symt2}\\
T^{\pi}_{-\lambda' -\lambda}\left( p',p,x' \right) = 4 \lambda' \lambda T^{\pi}_{\lambda' \lambda}\left( p',p,x' \right) . \label{symt3}
\end{eqnarray}
These symmetry relations in (\ref{symt1}), (\ref{symt2}), (\ref{symt3}) reduce the number of equations to be solved. For each parity state we need to solve the LS equation given in Eq.~(\ref{lse}) only for, say, $T^{\pi}_{\frac{1}{2} \frac{1}{2}}\left( p',p,x' \right)$ . 

\section{\label{example}Example of Calculations}

We show here just as an example some calculations for a simple spin-orbit potential given as 
\begin{equation}
V(r) = V_{c}(r) + V_{s}(r) \mathbf{l} \cdot \mathbf{s} , 
\end{equation}
with $\mathbf{l}$ being the angular momentum operator, $\mathbf{s} = \frac{1}{2} \ffat{\sigma}$, the radial functions $V_{c}(r)$ and $V_{s}(r)$ of the type of the Malfliet-Tjon\cite{mt} potential:
\begin{eqnarray}
V_{c}(r) = -V_{ca}\frac{e^{-\mu_{c}r}}{r} + V_{cb}\frac{e^{-2\mu_{c}r}}{r} , \qquad
V_{s}(r) = -V_{sa}\frac{e^{-\mu_{s}r}}{r} + V_{sb}\frac{e^{-2\mu_{s}r}}{r} , 
\end{eqnarray}
$V_{ca} = 3.22$, $V_{cb} = 7.39$, $\mu_{c} = 1.55$, $V_{sa} = 2.64$, $V_{sb} = 7.39$, $\mu_{s} = 0.63$. The mass of particle 1 acting as the projectile is of nucleon mass and that of particle 2 acting as the target is of kaon mass. It would certainly be much more interesting if we could test our work on KN scattering. But to do that we need a KN interaction in operator form, which is unfortunately not yet available. 

\begin{figure}[t!]
\centerline{\psfig{file=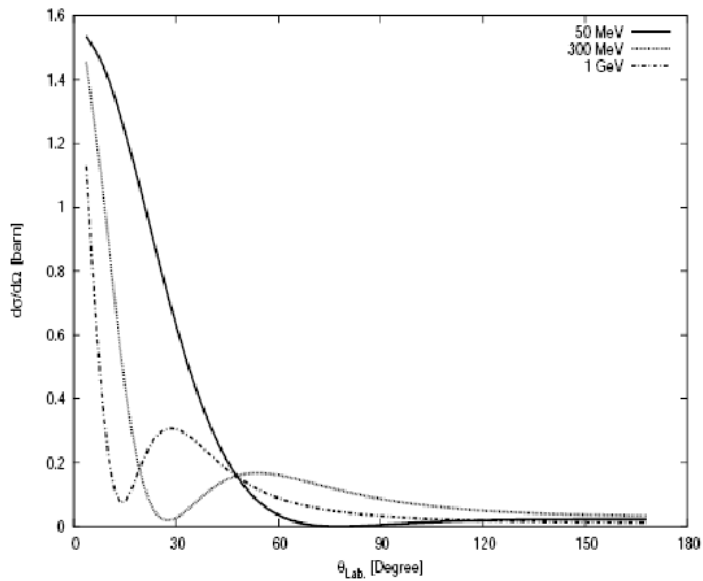,width=95mm}}
\vspace*{8pt}
\caption{The spin averaged differential cross section for various projectile’s laboratory energies\protect\label{observ}}
\end{figure}

Figure \ref{observ} shows the spin averaged differential cross section for various projectile’s laboratory energies from a few MeV up to 1 GeV. We want to stress out that the algebraic and numerical effort in 3D approach is the same for all different energies. 

\section{\label{summary}Summary}

We have developed a 3D technique for scattering of spin-$\frac{1}{2}$ and spin-$0$ particles. 
It will be very useful for investigation on, for example, KN system, given a KN interaction in 
operator form. The technique shows to be a good alternative to the PW technique, especially for 
higher energy region, where PW calculations may become not feasible. As an example some 
calculations based on a simple spin-orbit potential are performed.

\end{document}